\documentclass{book}
\usepackage[contrib,lang=british]{ems-book} 
\newtheorem{remark}{Remark}

\begin{document}
\mainmatter

\title{Derivation of macroscopic epidemic models from multi-agent systems}
\titlemark{Derivation of macroscopic epidemic models from multi-agent systems}

\emsauthor{1}{Mattia Zanella}{M.~Zanella}


\emsaffil{1}{Department of Mathematics "F. Casorati", University of Pavia, Via A. Ferrata 5, 27100 Pavia, Italy \email{mattia.zanella@unipv.it}}

\classification[92D30]{35Q84}



\begin{abstract}
We present a systematic review of some basic results on the derivation of classical epidemiological models from simple agent-based dynamics. The evolution of large populations is coupled with the dynamics of the contact distribution, providing insights into how individual behaviors impact macroscopic epidemiological trends. The resulting set of equations incorporates local characteristics of the operator governing the emergence of a family of contact distributions. To validate the proposed approach, we provide several numerical results based on asymptotic preserving methods, demonstrating their effectiveness in capturing the multi-scale nature of the problem and ensuring robust performance across different regimes.
\end{abstract}

\makecontribtitle


\section{Introduction}
In this work, we focus on deriving model-based transition rates for compartmental models in mathematical epidemiology from a multi-agent perspective. Among the key factors shaping the epidemic trajectory, the mixing patterns of a population have been widely investigated, particularly in relation to age distribution \cite{B_etal, Fumanelli, Mistry}. Special attention has been given to determining the impact of contact heterogeneities between individuals; see, for example, \cite{Ferguson, Germann, MA} and the references therein. The computation of human-mixing patterns has been approached through various methodologies, including surveys, contact diaries, sensors, and the observation of synthetic populations, as discussed in \cite{Catt, Dall, Mossong, Zha}. These methods provide a more detailed description of human behavior, which can be embedded in mathematical and computational analyses of the spread of infectious diseases. For instance, in \cite{APZ0, APZ1}, setting-specific (household, workplace, school) containment measures were proposed to selectively minimize the number of infections, taking into account the population's contact structure and potential heterogeneities.

The evolution of epidemics can be understood as the result of various interactions among a large number of agents from different species \cite{H}, whose transitions are influenced by multidimensional factors, including both behavioral aspects and the biological features of the pathogen. In this context, the legacy of kinetic theory emerges as a robust mathematical paradigm for analyzing the collective trends of large many-agent systems. Numerous research efforts have been dedicated to understanding the connection between microscopic agent-based dynamics and the evolution of observable data, with the goal of gaining a more detailed comprehension of these systems and their emergent behaviors \cite{BZ, CPS, De, DPTZ, LT, Z}. Indeed, kinetic equations bridge the gap between the microscopic scale of agents, whose dynamics are partially known and suffer of high variability, and the macroscopic observable scale, where data are typically available. The possibility to derive effective macroscopic descriptions that are consistent with microscopic interaction dynamics is of paramount importance to enhance to explanatory power of mathematical epidemiology and to comprehend the way in which the infection propagates in many-agent system with emerging features.  In particular, the design of effective public health interventions such as enhanced surveillance and social distancing should take into account community-specific strategies by leveraging microscopic properties of the contact dynamics. 

In this contribution, we introduce in Section \ref{sect:2} a stylized contact formation dynamics defining the multi-agent system. The evolution of observable quantities, together with emergence of equilibrium distribution, is obtained by means of a Fokker-Planck approach which can be derived for large systems. In Section \ref{sect:3}, we incorporate the contact dynamics in a classical SIR-type compartmental model for which the evolution of observable quantities are computed in a suitable limit. Finally, in Section \ref{sect:4} we present several numerical results based on structure preserving schemes to show consistency of the proposed approach.

\section{From agent-based to mean-field models for contact formation dynamics}
\label{sect:2}

We consider a large system of agents that are identified by their number of daily contacts $x_i(t) \in \mathbb R^+$, $i = 1,\dots,N$. The evolution of $x_i(t)$ can be obtained through several stochastic growth models having the following general structure
\begin{equation}
\label{eq:xi}
dx_i = B(x_i)\Psi(x_i)dt +\sqrt{\sigma^2 B(x_i)} x_i dW_i^t,
\end{equation}
being $\{W_i\}_{i=1}^N$ a set of independent Wiener processes weighted by the positive parameter $\sigma^2 >0$. In \eqref{eq:xi} two different mechanisms are responsible for the dynamics of $x_i(t)$ for all $i=1,\dots,N$:
\begin{itemize}
\item The drift function $\Psi(\cdot)$ which characterises deterministic variations of the contact structure due to environmental factors.  
\item The random process $W_i^t$, $i=1,\dots,N$, characterising the fluctuations in the contact variation due to unpredictable factors and to  population heterogeneities. 
\end{itemize}
The function $B(\cdot)>0$ is a kernel tuning the frequency of updates in relation to state $x_i$.  It is worth to remark that $\Psi(\cdot)$ may depend by additional features characterising the system of agents, like the age and other activity-specific quantities. For the sake of simplicity, in the current presentation we will assume that all the agents are indistinguishable and we point the interested reader to \cite{Z_etal} for an example including age of participants. 

Amongst the most common examples for the drift function $\Psi(\cdot)$, we find the logistic growth model, which is defined as follows
\begin{equation}
\label{eq:logistic_xi}
\Psi(x_i) = \dfrac{\alpha}{2} x_i\left(1-\dfrac{x_i}{m} \right), 
\end{equation}
where $\alpha>0$ is the growth rate and $m>0$ is the so-called carrying capacity of the system of agents and may depend on environmental factors. Logistic-type growths are often considered in ecology and in the analysis of socio-economic systems due to their effectivity in describing a self-limiting growths of a population due to capacity constraints, see e.g. \cite{HP,JMS,K}. 

Another relevant example is provided by the von Bertalanffy growth model \cite{vonB}. This model can be  postulated through energy conservation principles and is provided by the following choice
\begin{equation}
\label{eq:bertalanffy_xi}
\Psi(x_i) = \dfrac{\alpha}{2} x_i \left( x_i^{a-1} - q \right), \qquad q  = m^{a-1},
\end{equation}
being $0 \le a <1$ and $\alpha >0$ is the growth rate. Recently, the von Bertalanffy growth model has been considered to describe ontogenic growth phenomena, see \cite{WBE}. The last example that we will discuss is provided by the so-called Gompertz growth model having the following form
\begin{equation}
\label{eq:gompertz_xi}
\Psi(x_i) = \dfrac{\alpha}{2} x_i \log \left( \dfrac{m}{x_i}\right), 
\end{equation}
Typical applications of the Gompertz-type growth model can be found in tumour growth dynamics \cite{G,RBKW}.  As observed in \cite{TPZ} all these models can be framed in a generalised growth model defined by a transition function  
\begin{equation}
\label{eq:general_xi}
\Psi(x_i) = \dfrac{\alpha}{2\delta}x_i  \left( 1- \left(\dfrac{x_i}{m}\right)^{\delta} \right), 
\end{equation}
parametrised by $\alpha>0$, $\delta \in [-1,1]$ and having carrying capacity $m>0$. We can observe that  \eqref{eq:general_xi} generates a logistic-type model for any $\delta>0$, a von Bertalaffy-type model in the case $\delta<0$, whereas the Gompertz growth model is obtained in the limit $\delta\to 0$. In all the considered examples, setting $\sigma^2\equiv 0$ we can explicitly show that $x_i\to m$ exponentially in time, meaning that all the agents are trying to reach asymptotically a given number of contacts. 

In \cite{B_etal, Mossong}, it has been shown that mixing patterns were very similar across the considered countries and that people of the same age tended to mix with each other, having a country-specific mean number of daily contacts. Additional information was extracted from the contact distribution, which is, as before, country-specific and Gamma-like in shape. In the following, we will show how similar information can be recovered by examining the case of a large number of agents.

Following \cite{TPZ} we can show that when $N \to +\infty$ the behavior of the system can be expressed in terms of a Fokker-Planck equation for the evolution of the probability density $f(x,t)$ associated to a stochastic process $X(t) \sim f(x,t)$ which gives the statistical distribution of the number of daily contacts in a certain group of agents. The quantity $f(x,t)dx$ represents the fraction of agents which, at time $t\ge0$, are characterised by a number of daily contacts in the interval $[x,x+dx)$. The knowledge of the dynamics of the probability density $f(x,t)$ allows to observe the emergence of stylized facts and the formation of patterns. The evolution of $f(x,t)$, $   x \in \mathbb R^+$, is provided by the Fokker-Planck-type equation
\begin{equation}
\label{eq:FP}
\begin{split}
\partial_t f(x,t)  = \partial_x \left[ \dfrac{\alpha}{2\delta} \left( \left(\dfrac{x}{m} \right)^\delta-1\right) B(x)x f(x,t) + \dfrac{\sigma^2}{2}\partial_x (B(x)x^2 f(x,t)) \right],
\end{split}
\end{equation}
where $B(x)>0$ is an interaction kernel tuning the frequency of interactions. The equation \eqref{eq:FP} is coupled with no-flux boundary conditions
\begin{equation}
\label{eq:noflux}
\begin{split}
 \dfrac{\alpha}{2\delta} \left( \left(\dfrac{x}{m} \right)^\delta-1\right) B(x)x f(x,t) + \dfrac{\sigma^2}{2}\partial_x (B(x)x^2 f(x,t))\Big|_{x = 0} = 0, \\
 x^2 f(x,t)\Big|_{x=0} = 0. 
\end{split}
\end{equation}
A rigorous derivation of such partial differential equation can be obtained  through classical results of stochastic differential equations \cite{LeBL,Pav, Risken} or via a kinetic-type approach as presented in \cite{DPTZ,FMZ}. Positivity preservation and an $L^1$ contraction property for the equation \eqref{eq:FP} have been studied in \cite{FMZ}. 

Proceeding as in \cite{FMZ}, we may notice that the mass is always conserved in \eqref{eq:FP}, whereas the mean becomes a conserved quantity provided that  $B(x) = x^{-\frac{1+\delta}{2}}$. More specifically, we have
\begin{equation}
\dfrac{d}{dt} \int_{\mathbb R_+}x f(x,t)dx = -\dfrac{\alpha}{2\delta}\int_{\mathbb R_+} \left( \left(\dfrac{x}{m} \right)^\delta-1\right)x^{1-\frac{1+\delta}{2}}f(x,t)dx = 0
\end{equation}
provided $\delta = \pm 1$.  

\begin{remark}
In \cite{DPTZ} it has been considered an alternative approach which exploits elementary variation having the form 
\begin{equation}
\label{eq:micro_kin}
x^\prime = x + \Phi_\epsilon^\delta(x/m) + x \eta_\epsilon, 
\end{equation}
with 
\[
\Phi_\epsilon^\delta(s) = \alpha \dfrac{1-e^{\epsilon(s^\delta -1)/\delta}}{e^{\epsilon (s^\delta-1)/\delta}+1}, 
\]
and $\eta_\epsilon$ is a centered random variable with finite variance $\left\langle\eta_\epsilon^2 \right\rangle = \epsilon\sigma^2$. We remark that for $\epsilon\ll 1$ the following approximation holds
$$\Phi_\epsilon^\delta \approx \epsilon \dfrac{\alpha}{2\delta} \left(\left( \dfrac{x}{m}\right)^\delta-1 \right). $$
 Hence, the introduced $\Phi_\epsilon^\delta(\cdot)$ has the structure of the drift function \eqref{eq:general_xi} introduced in the agent-based stochastic dynamics \eqref{eq:xi}. If $g_\epsilon(x,t)dx$ is the fraction of agents with a number of contact between $x$ and $x+dx$,  the time evolution of the distribution $g_\epsilon(x,t)$ satisfies is therefore given by a Boltzmann-type equation linking the particles' dynamics with their aggregate trends. In weak form, for any test function $\varphi(x)$, we have
\begin{equation}
\label{eq:Boltz}
\partial_t \int_{\mathbb R_+ } \varphi(x)g_\epsilon(x,t)dx = \int_{\mathbb R_+} B(x) \left\langle \varphi(x^\prime) - \varphi(x) \right\rangle g_\epsilon(x,t)dx. 
\end{equation}
The kinetic model \eqref{eq:Boltz} conserves the same quantities of \eqref{eq:FP}. On the other hand, we highlight that the large time behaviour of \eqref{eq:Boltz} is very difficult to compute explicitly, see \cite{PT}. As shown in \cite{DPTZ}, for $\epsilon\to 0^+$, the solution of the Boltzmann-type model \eqref{eq:Boltz} converges, up to the extraction of a subsequence, to a probability density $f(x,t)$, which is a weak solution of the Fokker-Planck equation \eqref{eq:FP}. 
\end{remark}

\subsection{Equilibrium states}\label{sect:equi}
In the following we concentrate on the equilibrium distribution of the Fokker-Planck model \eqref{eq:FP}. The large time distribution $f^\infty(x)$ is the solution of the differential equation 
\[
\partial_x (x^{2-\frac{1+\delta}{2}} f^\infty(x))  + \dfrac{\gamma}{\delta} \left(\left( \dfrac{x}{m}\right)^\delta - 1 \right)x^{1-\frac{1+\delta}{2}}f^\infty(x)=0, \qquad \gamma = \alpha/\sigma^2,
\]
which is given by 
\begin{equation}
\label{eq:finfty_general}
f^\infty(x)  = C_{m,\delta,\gamma,\sigma^2} x^{\frac{\gamma}{\delta}-2+\frac{1+\delta}{2}}\exp\left\{ - \dfrac{\gamma}{\delta^2} \left(\dfrac{x}{m}\right)^\delta \right\}
\end{equation}
with $C_{m,\delta,\gamma,\sigma^2}>0$ a normalization constant. 
More in detail, in the case $\delta>0$, corresponding to logistic growths as presented in \eqref{eq:logistic_xi}, the probability density $f^\infty(x)$ is a generalized Gamma density, which is characterised in terms of a shape parameter $\kappa>0$, a scale parameter $\theta>0$ and the exponent $\delta>0$ and reads
\begin{equation}
\label{eq:finfty_deltapos}
f^\infty_{\kappa,\delta,\theta,m}(x) = \dfrac{\delta}{\theta^\kappa \Gamma(\kappa)} x^{\kappa-1}\exp\left\{ - (x/\theta)^\delta\right\},
\end{equation}
where 
\[
\kappa = \dfrac{\gamma}{\delta} + \dfrac{1+\delta}{2}-1,\qquad \theta = m \left(\dfrac{\delta^2}{\gamma}\right)^{1/\delta}. 
\]
The relaxation of the Fokker-Planck-type equation \eqref{eq:FP} to the generalized Gamma equilibrium \eqref{eq:finfty_deltapos} has been studied  in \cite{Toscani_Gamma}. The emerging generalized Gamma equilibra are characterised by slim tails decaying to zero exponentially as $x \to +\infty$. In the case $\delta\equiv 1$ we can observe that \eqref{eq:finfty_deltapos} is a Gamma density for which the mean and energy are explicitly given by 
\begin{equation}
\label{eq:mean_var_gamma}
\int_{\mathbb R_+}xf^\infty_{\kappa,\delta,\theta,m}(x) dx = m\qquad \int_{\mathbb R_+}x^2 f^\infty_{\kappa,\delta,\theta,m}(x)dx = \dfrac{\gamma+1}{\gamma}m^2. 
\end{equation}

If $\delta<0$, corresponding to the von Bertalanffy growth defined in \eqref{eq:bertalanffy_xi} the probability density $f^\infty(x)$ is a power-law distribution having the form 
\begin{equation}
\label{eq:finfty_deltaneg}
f^\infty_{\kappa,\delta,\theta,m}(x) = \dfrac{|\delta|}{\Gamma(\kappa/|\delta|)}\dfrac{\theta^\kappa}{x^{\kappa+1}}\exp\left\{- \left(\dfrac{\theta}{x} \right)^{|\delta|}\right\},
\end{equation}
with 
\[
\kappa = \dfrac{\gamma}{|\delta|}+\dfrac{1+\delta}{2}+1, \qquad \theta = m \left( \dfrac{\gamma}{\delta^2}\right)^{1/|\delta|}, 
\]
being as before $\gamma = \alpha/\sigma^2>0$. 
Therefore, we may notice that, if $\delta<0$, the emerging  equilibrium 
has polynomial decay for $x \to +\infty$ and has bounded moments of order $p<\kappa$. In \eqref{eq:finfty_deltaneg} the case $\delta \equiv -1$ corresponds to the inverse Gamma distribution, we refer the interested reader to \cite{FPTT} for more details insight on the convergence properties of kinetic-type equations towards fat-tailed distributions. In this case, mean and energy are given by 
\begin{equation}
\label{eq:mean_var_invgamma}
\int_{\mathbb R_+}x f^\infty_{\kappa,\delta,\theta,m}(x)dx = m, \qquad \int_{\mathbb R_+}x^2 f^\infty_{\kappa,\delta,\theta,m}(x) dx =  \dfrac{\gamma}{\gamma-1}m^2, 
\end{equation}
with $\gamma = \alpha/\sigma^2$, $\alpha\neq \sigma^2$. 
Hence, the parameter $\delta \in [-1,1]$ is able to describe societies that are characterized by a radically different contact distribution in terms of the tail behaviour. 

\begin{figure}
\centering
\includegraphics[scale = 0.3]{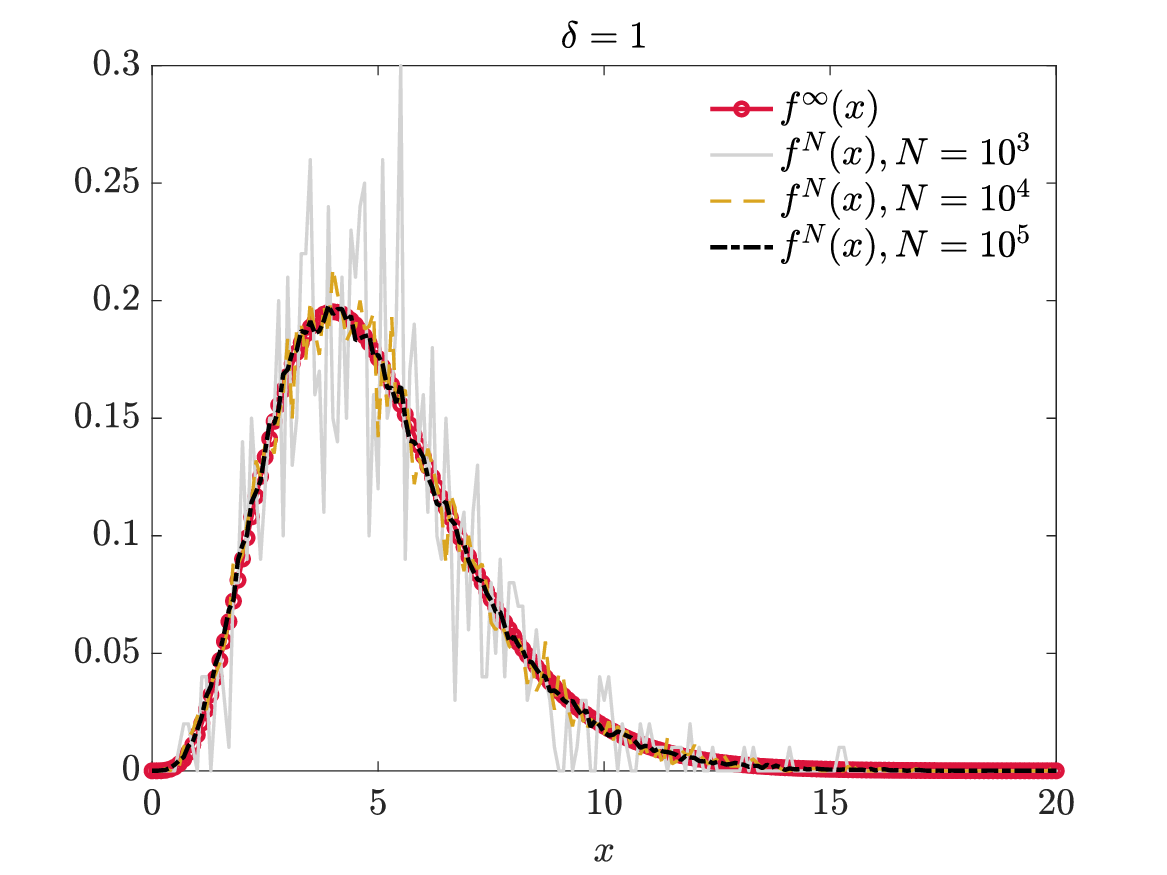}
\includegraphics[scale = 0.3]{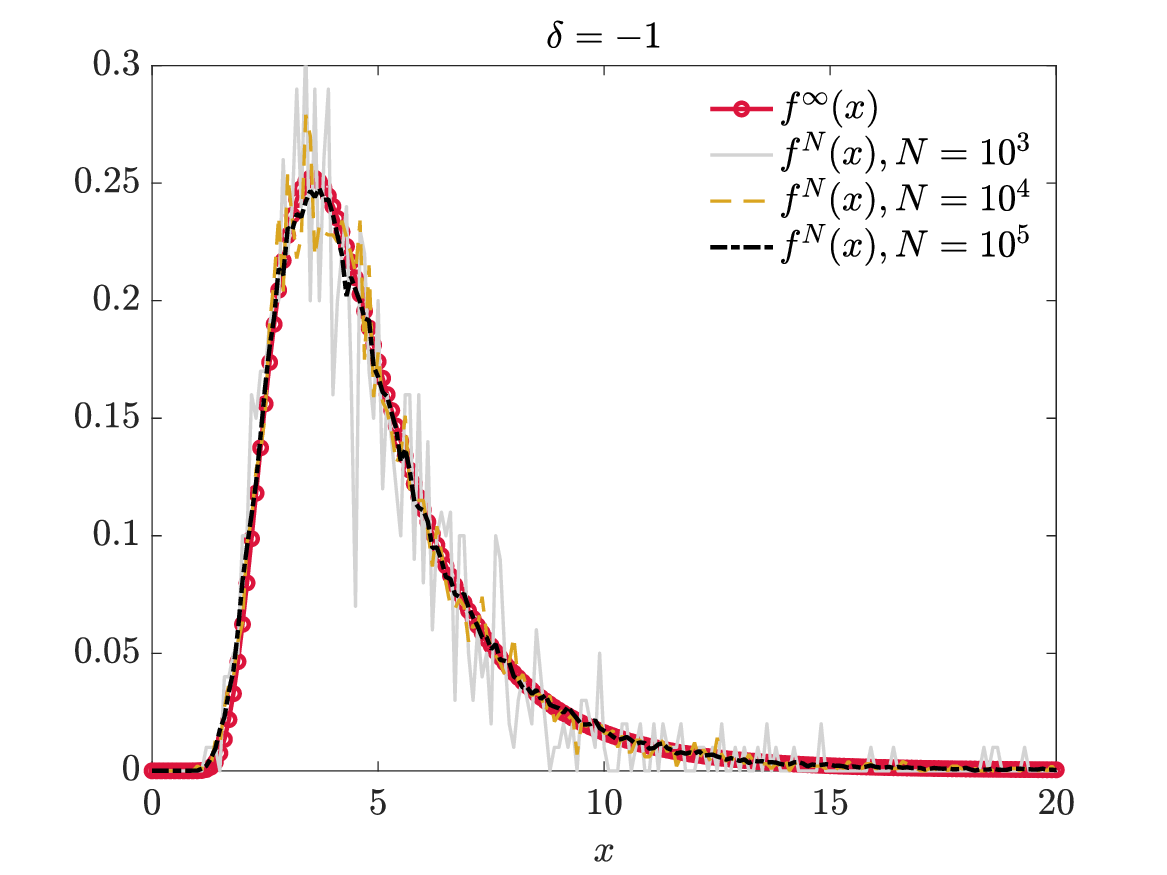}\\
\includegraphics[scale = 0.3]{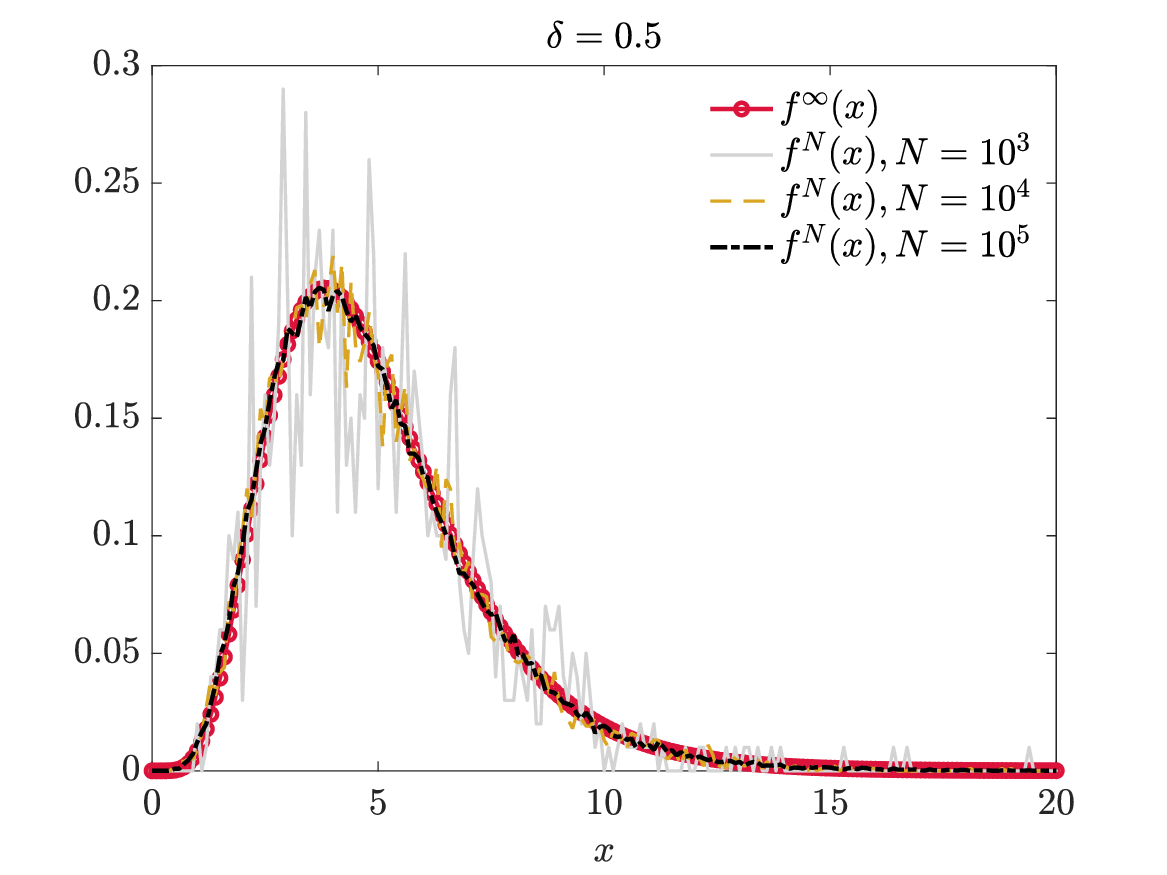}
\includegraphics[scale = 0.3]{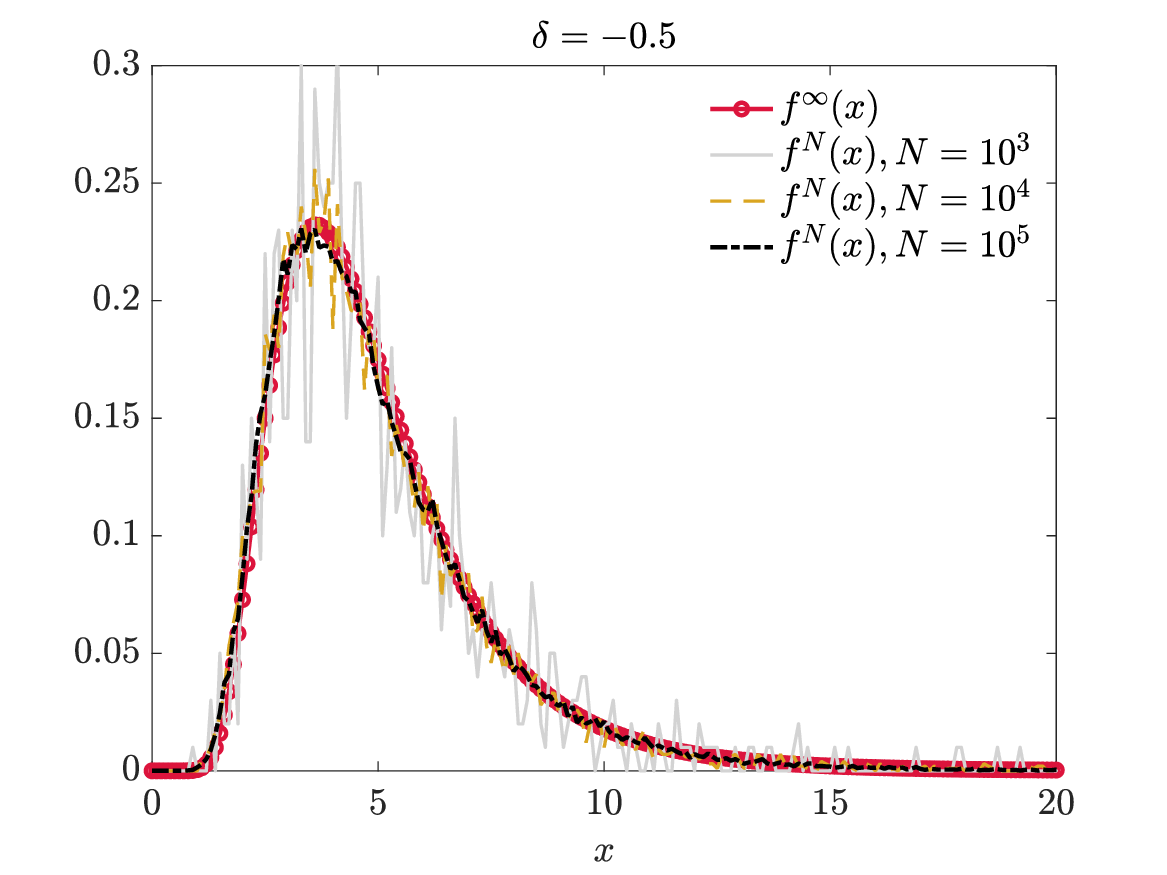}
\caption{Comparison of the emerging empirical equilibrium distribution for the system of agent $\{x_i\}_{i=1}$ whose dynamics is given in \eqref{eq:xi} with transition \eqref{eq:general_xi} over the time interval $[0,100]$. We considered an Euler-Maruyama approximation of the dynamics with $\Delta t = 10^{-2}$. We considered an artificial choice of the parameters given by $\alpha = 1$, $\sigma^2=0.2$, $m = 9$ and several choices for the parameter $\delta = \pm 0.5,\pm 1$. }
\label{fig:part}
\end{figure}

In order to compare the emerging distribution $f^\infty_{\kappa,\delta,\theta,m}(x)$ with the one obtained from a large system of particles $\{x_i\}_{i=1}^N$ we compute the empirical density 
\[
f^N(x,t) = \dfrac{1}{N} \sum_{i=1}^N \delta(x-x_i(t)). 
\]
In Figure \ref{fig:part} we compare $f^\infty_{\kappa,\delta,\theta,m}(x)$ with the large time distribution $f^N$ for an increasing number of agents $N = 10^3,10^4,10^5$. The empirical density has been mollified through the convolution with a normal density. The evolution of the particle dynamics is given by \eqref{eq:xi} with $\Psi(\cdot)$ as in \eqref{eq:general_xi} and a kernel $B(x_i) = x_i^{-\frac{1+\delta}{2}}$. At the numerical level we considered an Euler-Maruyama method over the time interval $[0,T]$, $T=100$, with $\Delta t = 10^{-2}$. As initial condition for the particle system we considered a sample of the uniform distribution  
\[
f_0(x) = 
\begin{cases}
\dfrac{1}{2} & x \in [4,6] \\
0 & x \in \mathbb R \setminus [4,6]. 
\end{cases}
\]
In the test we have considered $m = \int_{\mathbb R_+}xf_0(x)dx$. We can observe how the large time behaviour of the system of agents is in agreement with the theoretical equilibrium distribution obtained from the Fokker-Planck equation. 

\begin{remark}
We highlight how the Fokker-Planck equation \eqref{eq:FP} can be seen as a reduced complexity model with respect to a Boltzmann-type equation since it allows to compute the emerging equilibrium distribution of the system of agents. 
\end{remark}

\section{Kinetic compartmental model and observable dynamics}\label{sect:3}
In this section, we introduce a compartmental model that for the coupled evolution of the epidemic in a large system of individuals coupled with a contact formation dynamics. In more detail, we consider a system of agents subdivided into the following epidemiologically relevant states: susceptible ($S$) agents are the ones that can contract the disease, infected infectious ($I$) agents are responsible for the spread of the disease, and removed ($R$) agents cannot spread the disease.   To incorporate the impact of contact distribution in the dynamics we consider the densities $f_J(x,t)$, $J \in\mathcal C =  \{S,I,R\}$ such that 
\[
\sum_{J \in \mathcal C}f_J(x,t) = f(x,t), \qquad \int_{\mathbb R_+}f(x,t)dx = 1. 
\]
Hence, the mass fractions of the population in each compartment is defined as follows
\[
\rho_J(t) = \int_{\mathbb R_+}f_J(x,t), 
\]
and their moments of order $r>0$ are given by 
\[
\rho_J(t)m_{r,J}(t) = \int_{\mathbb R_+}x^r f_J(x,t)dx. 
\]

The kinetic model defining the time evolution of $f_J(x,t)$ follows by combining the epidemic process with the contact formation dynamics
\begin{equation}
\label{eq:kineticSIR}
\begin{split}
\partial_t f_S(x,t) &= -K(f_S,f_I)(x,t) + \dfrac{1}{\tau}Q_S(f_S)(x,t), \\
\partial_t f_I(x,t) &=K(f_S,f_I)(x,t) - \gamma_I f_I(x,t) + \dfrac{1}{\tau}Q_I(f_I)(x,t), \\
\partial_t f_R(x,t) &=\gamma_I f_I(x,t) + \dfrac{1}{\tau}Q_R(f_R)(x,t).
\end{split}
\end{equation}
In \eqref{eq:kineticSIR} the operators $Q_J(f_J)(\cdot,\cdot)$ characterize the emergence of a contact distribution. In the following, we will concentrate on the family of distributions defined in Section \ref{sect:equi}. The transmission of the infection is governed by the local incidence rate
\begin{equation}
\label{eq:K_spreading}
K(f_S,f_I)(x,t) = f_S(x,t) \int_{\mathbb R_+}\kappa(x,x_*) f_I(x_*,t)dx_*,
\end{equation}
being $\kappa(\cdot,\cdot)$ a nonnegative contact function measuring the impact of contact rates among different compartments. Following \cite{DPTZ,DTZ} we concentrate on the case
\begin{equation}
\label{eq:kappa}
\kappa(x,x_*) = \beta x^c x_*^c,
\end{equation}
with $c\ge0$ and $\beta>0$. The parameter $\gamma_I>0$ is the recovery rate and the relaxation parameter $\tau>0$ determines the relaxation scale of the contact distribution of the agents towards an equilibrium in response to the epidemic dynamics. In the following, we will consider only the case $c=1$ for the sake of simplicity, we point the interested reader to \cite{MZ} for the case of a general $c>0$. 

We recall that the operators $Q_J(\cdot)$ are therefore chosen according to \eqref{eq:FP} and have the following form 
\begin{equation}
\label{eq:QJ}
Q_J(f_J) = \partial_x \left[\dfrac{\alpha}{2\delta} \left( \left(\dfrac{x}{m} \right)^\delta-1\right)x^{\frac{1-\delta}{2}}f_J(x,t) + \dfrac{\sigma^2}{2} \partial_x (x^2 f_J(x,t)) \right]
\end{equation}
coupled with no-flux boundary conditions \eqref{eq:noflux}. Positivity and uniqueness of the solutions for the system of kinetic equations \eqref{eq:kineticSIR} have been discussed in \cite{FMZ}. 

\begin{remark}
The kinetic model \eqref{eq:kineticSIR} is based on a SIR compartmentalization. Nevertheless, it is possible to adapt the proposed method to different compartmentalizations of the society. We refer the interested reader to \cite{Z,Z_etal} for other examples involving different compartments. 
\end{remark}

\subsection{Macroscopic modelling}

In this section, we derive the evolution of observable quantities of the kinetic system of equations \eqref{eq:kineticSIR}. In Section \ref{sect:2} we have observed how the invariant quantities of the contact formation dynamics are given by mass and momentum, for which we wish to obtain a closed system of equation describing their evolution.

It is worth noticing that if $c = 0$ in \eqref{eq:kappa}, meaning that there is no influence of the contact distribution in the infection dynamics, the evolution of the mass fractions follows is given by 
\begin{equation}
\begin{split}
\dfrac{d}{dt}\rho_S&= -\beta \rho_S\rho_I \\
\dfrac{d}{dt}\rho_I&= \beta \rho_S \rho_I - \gamma_I \rho_I \\
\dfrac{d}{dt} \rho_R&= \gamma_I \rho_I,
\end{split}
\end{equation}
corresponding to the classical SIR model which indeed is based on the assumption of homogeneity of the population. The dynamics of  the mean number of daily contacts is obtained by computing $\int_{\mathbb R_+}xf_J(x,t)dx$, $J \in \{S,I,R\}$, from \eqref{eq:kineticSIR} and results constant in time, i.e.
\begin{equation}
\begin{split}
\dfrac{d}{dt}(\rho_S m_S)&=-\beta \rho_S \rho_Im_I\\
\dfrac{d}{dt}(\rho_I m_I )&= \beta \rho_S\rho_Im_I - \gamma_I \rho_I m_I\\
\dfrac{d}{dt}(\rho_R m_R) &= \gamma_I \rho_I m_I , 
\end{split}
\end{equation}
from which we deduce that $\frac{d}{dt} m_J = 0$ for $J \in \mathcal C$. Therefore,  in the homogeneity assumption there is no change in the contact structure due to the epidemic spreading since all the agents become infected with the same rate. 

On the other hand, wishing to incorporate an influence of the number of daily contacts on the epidemic transmission we present here the prototypical case where we fix $c=1$ in \eqref{eq:K_spreading}. In this case, the evolution of the mass fractions is given by 
\begin{equation}
\begin{split}
\label{eq:macro_mass}
\dfrac{d}{dt} \rho_S &= -\beta \rho_S m_S \rho_I m_I \\
\dfrac{d}{dt} \rho_I &= \beta \rho_S m_S \rho_I m_I - \gamma_I \rho_I \\
\dfrac{d}{dt} \rho_R &= \gamma_I \rho_I, 
\end{split}
\end{equation}
where the obtained system of equations is coupled with the evolution of the momentum in each compartment $J \in \mathcal C$
\begin{equation}
\begin{split}
\dfrac{d}{dt} (\rho_S m_S)& = -\beta \int_{\mathbb R_+}x^2 f_S(x,t)dx \rho_I m_I \\
\dfrac{d}{dt} (\rho_I m_I )&=  \beta \int_{\mathbb R_+}x^2 f_S(x,t)dx  \rho_I m_I - \gamma_I \rho_I \\
\dfrac{d}{dt} (\rho_R m_R) &= \gamma_I \rho_I. 
\end{split}
\end{equation}
The obtained system of equations is therefore not closed since it depends on the kinetic density. The idea behind \cite{DPTZ} is to exploit the two scale structure of \eqref{eq:kineticSIR}  such that if $\tau\ll 1$ the densities $f_J(x,t)$ converge towards their equilibrium parametrised by mass and momentum. This assumption  is rather classical in contact formation dynamics and is frequently observed in empirical data-oriented studies \cite{Zha}. We further highlight how behavioural aspects are frequently considered very fast with respect to the evolution of an epidemic, see e.g.  \cite{Pol}. 

 Hence, we may introduce the following approximation which is reminiscent of the equilibrium closure in classical kinetic theory, see e.g. \cite{BGL,BCGP,Cerc}. In the limit $\tau\to 0^+$, we may consider $f_J(x,t)$ close to the local equilibrium density and we get
\[
\int_{\mathbb R_+} x^2 f_J(x,t)dx \approx \int_{\mathbb R_+}x^2 \rho_J(t) f_{\kappa,\delta,\theta,m_J}(x,t)dx, 
\]
where $f_{\kappa,\delta,\theta,m_J}(x,t)$ has been defined in \eqref{eq:finfty_deltapos} and in \eqref{eq:finfty_deltaneg}. Therefore, we remark how the evolution of the mean number of contact is influenced by the equilibrium density coming from the logistic-type contact formation dynamics defined, at the particle level, by \eqref{eq:xi} with transition function \eqref{eq:logistic_xi}. We first  concentrate on the relevant case $\delta\equiv 1$ leading to the Gamma density \eqref{eq:finfty_deltapos} for which the relation between mean and energy is given by \eqref{eq:mean_var_gamma}. Hence, in the case $\delta\equiv 1$ the system for the evolution of the mean contacts reads 
\begin{equation}
\begin{split}
\dfrac{d}{dt}( \rho_Sm_S) &=-\beta \rho_S\rho_I m_I \dfrac{\gamma+1}{\gamma}m_S^2 \\
\dfrac{d}{dt} (\rho_Im_I)&= \beta \rho_S\rho_I m_I \dfrac{\gamma+1}{\gamma}m_S^2  -\gamma_I \rho_I m_I \\
\dfrac{d}{dt}( \rho_Rm_R) &= \gamma_I \rho_Im_I, 
\end{split}
\end{equation}
being $\gamma = \alpha/\sigma^2>0$, which gives
\begin{equation}
\begin{split}
\label{eq:macro_mean_deltapos}
\dfrac{d}{dt}m_S &= -\dfrac{\beta}{\gamma}\rho_I m_I m_S^2 \\
\dfrac{d}{dt}m_I &= \beta \rho_S m_S m_I\left( \dfrac{\gamma+1}{\gamma}m_S-m_I \right) \\
\dfrac{d}{dt}m_R &= \gamma_I\dfrac{\rho_I}{\rho_R} \left(m_I -  m_R \right)
\end{split}
\end{equation}
The system of equations is therefore given by the coupled system for mass fractions and mean number of contacts \eqref{eq:macro_mass}-\eqref{eq:macro_mean_deltapos}. 

On the other hand, if we consider the case $\delta = -1$ we focus on a population with an emerging inverse Gamma-type contact distribution, which is characterised by an overpopulated power-law-type tail. In this case, the relation between mean and energy is given by \eqref{eq:mean_var_invgamma}  and  the macroscopic equation for the evolution fo the mean number of contacts is given by 
\begin{equation}
\begin{split}
\dfrac{d}{dt}( \rho_Sm_S) &=-\beta \rho_S\rho_I m_I \dfrac{\gamma}{\gamma-1}m_S^2 \\
\dfrac{d}{dt} (\rho_Im_I)&= \beta \rho_S\rho_I m_I \dfrac{\gamma}{\gamma-1}m_S^2  -\gamma_I \rho_I m_I \\
\dfrac{d}{dt}( \rho_Rm_R) &= \gamma_I \rho_Im_I, 
\end{split}
\end{equation}
which gives 
\begin{equation}
\begin{split}
\label{eq:macro_mean_deltaneg}
\dfrac{d}{dt}m_S &= -\dfrac{\beta}{\gamma-1}\rho_I m_I m_S^2 \\
\dfrac{d}{dt}m_I &= \beta \rho_S m_S m_I \left(\dfrac{\gamma}{\gamma-1}m_S - m_I \right) \\
\dfrac{d}{dt}m_R &= \gamma_I \dfrac{\rho_I }{\rho_R}\left( m_I-m_R\right). 
\end{split}
\end{equation}
The system of equations is therefore given by the coupled system for mass fractions and mean number of contacts \eqref{eq:macro_mass}-\eqref{eq:macro_mean_deltaneg}. 

The obtained macroscopic systems are not equivalent and they strongly depend on the contact formation dynamics of the population of interest. 
\begin{remark}
In general, we have limited information on the contact structure of a population and the propagation of uncertain quantities should be taken into account.   Furthermore, the construction of robust control methods for these models for the minimization of the number of infected should incorporate the heterogeneous contact structure of a population to guarantee the effectiveness of the considered strategy. A first approach has been proposed in \cite{FMZ}. 
\end{remark}

\section{Numerical tests}\label{sect:4}

In this section we present several numerical examples to show the consistency of the proposed approach. To numerically approximate the dynamics of the kinetic SIR model \eqref{eq:kineticSIR} for small values of $\tau\ll 1$, we will resort to strong stability preserving schemes for the epidemic dynamics combined to an implicit structure preserving scheme for the relaxation dynamics based on Fokker-Planck equations \cite{PZ}. These methods are capable of reproducing large time statistical properties of the exact steady state with arbitrary accuracy together with the preservation of the main physical properties of the solution, like positivity, large time behaviour and entropy dissipation. 

We compare the evolution of mass and local mean of the distribution $f_J(x,t)$, $J \in \mathcal C$, with the evolution of the obtained macroscopic systems, either of the form \eqref{eq:macro_mass}-\eqref{eq:macro_mean_deltapos} in the case $\delta \equiv 1$, or of the form \eqref{eq:macro_mass}-\eqref{eq:macro_mean_deltaneg} in the case $\delta\equiv -1$. 

Hence, we are interested in the evolution of $f_J(x,t)$, $J \in \mathcal C$, $x \in \mathbb R_+$, $t\ge0$, solution to \eqref{eq:kineticSIR}  and complemented by the initial condition $f_J(x,0) = f_J^0$. We consider a time discretization of the interval $[0,t_{\textrm{max}}]$ of size $\Delta t>0$. We denote by $f_J^n(x)$ the approximation of $f_J(x,t^n)$. We introduce a splitting strategy between the Fokker-Planck step $f_J^* = \mathcal O_{\Delta t}(f_J^n)$
\begin{equation}
\begin{cases}\label{eq:collision}
\partial_t f_J^* = \dfrac{1}{\tau}Q_J(f_J), \\
f_J^*(x,0) = f_J^n(x), \qquad J \in \mathcal C,
\end{cases}
\end{equation}
and the epidemiological step $f_J^{**} = \mathcal E_{\Delta t}(f_J^{**})$
\begin{equation}
\begin{cases}\label{eq:reaction}
\partial_t f_S^{**} = - K(f_S^{**},f_I^{**})\\
\partial_t f_I^{**} = K(f_S^{**},f_I^{**}) - \gamma_I f_I^{**} \\
\partial_t f_R^{**}= \gamma_I f_I^{**}, \\
f_J^{**}(x,0) = f_J^{*}(x,\Delta t), \qquad J \in \mathcal C. 
\end{cases}
\end{equation}
The operators $Q_J(\cdot)(x,t)$ have been defined in \eqref{eq:QJ} and are coupled with no-flux boundary conditions. Hence, the solution at time $t^{n+1}$ is given by the combination of the two introduced steps \eqref{eq:collision}-\eqref{eq:reaction}. A first-order splitting strategy corresponds to 
\[
f^{n+1}_J(x) = \mathcal E_{\Delta t}(\mathcal O_{\Delta t}(f_J^n(x))), 
\]
whereas the second-order Strang splitting method is obtained as 
\[
f_J^{n+1}(x) = \mathcal E_{\Delta t/2}(\mathcal O_{\Delta t}(\mathcal E_{\Delta t/2}(f_J^n(x)))), \qquad J \in \mathcal C. 
\]
The thermalization towards the contact distribution is solved by means of an implicit structure-preserving (SP) scheme for Fokker-Planck equations \cite{PZ}. The integration of the epidemiological step \eqref{eq:reaction} is performed through an RK4 method. 

\subsection{Consistency of macroscopic modelling via Gamma closure approximation }\label{sect:cons_pos}

We compare the evolution of observable quantities 
\begin{equation}
\label{eq:obs_num}
\rho_J^\tau(t) = \int_{\mathbb R_+}f_J(x,t)dx, \qquad m_J^\tau(t) = \dfrac{1}{\rho_J^\tau}\int_{\mathbb R_+}xf_J(x,t)dx, 
\end{equation}
where $f_J(x,t)$ is solution to \eqref{eq:kineticSIR} with $\delta = 1$, with $\rho_J(t)$, $m_J(t)$ characterising the macroscopic model obtained with a Gamma closure approximation, see  \eqref{eq:macro_mass}-\eqref{eq:macro_mean_deltapos}. Since  We approximate numerically \eqref{eq:kineticSIR} with mean-field contact dynamics operator $Q_J(\cdot)$, $J \in \mathcal C$, defined in \eqref{eq:QJ}. We introduce a uniform grid of the interval $[0,L]$ characterised by $\Delta x>0$ and we introduce the time discretization such that $t^n = n\Delta t$, $\Delta t>0$ and $n = 0,\dots,T$ with $T\Delta t = t_{\textrm{max}} = 50$. We point the interested reader to \cite{PZ} for the details on the adopted numerical scheme. In the following we considered $L = 50$ and $\Delta x  = \frac{1}{10}$. We considered as initial distribution 
\begin{equation}
\label{eq:init_dist}
f(x,0) = 
\begin{cases}
\dfrac{1}{4} & x \in [8,12] \\
0 & x \in \mathbb R\setminus [8,12]. 
\end{cases}
\end{equation}

\begin{figure}
\centering
\includegraphics[scale  = 0.20]{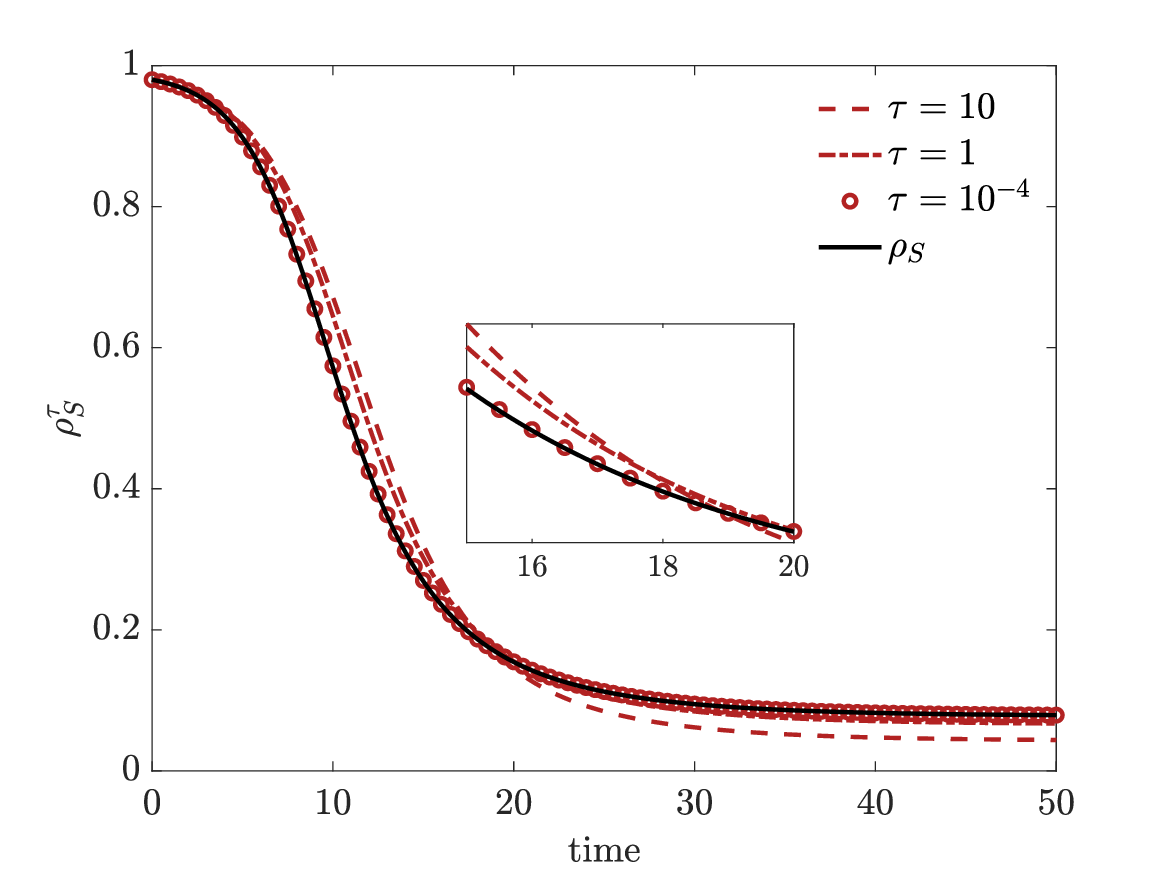}
\includegraphics[scale  = 0.20]{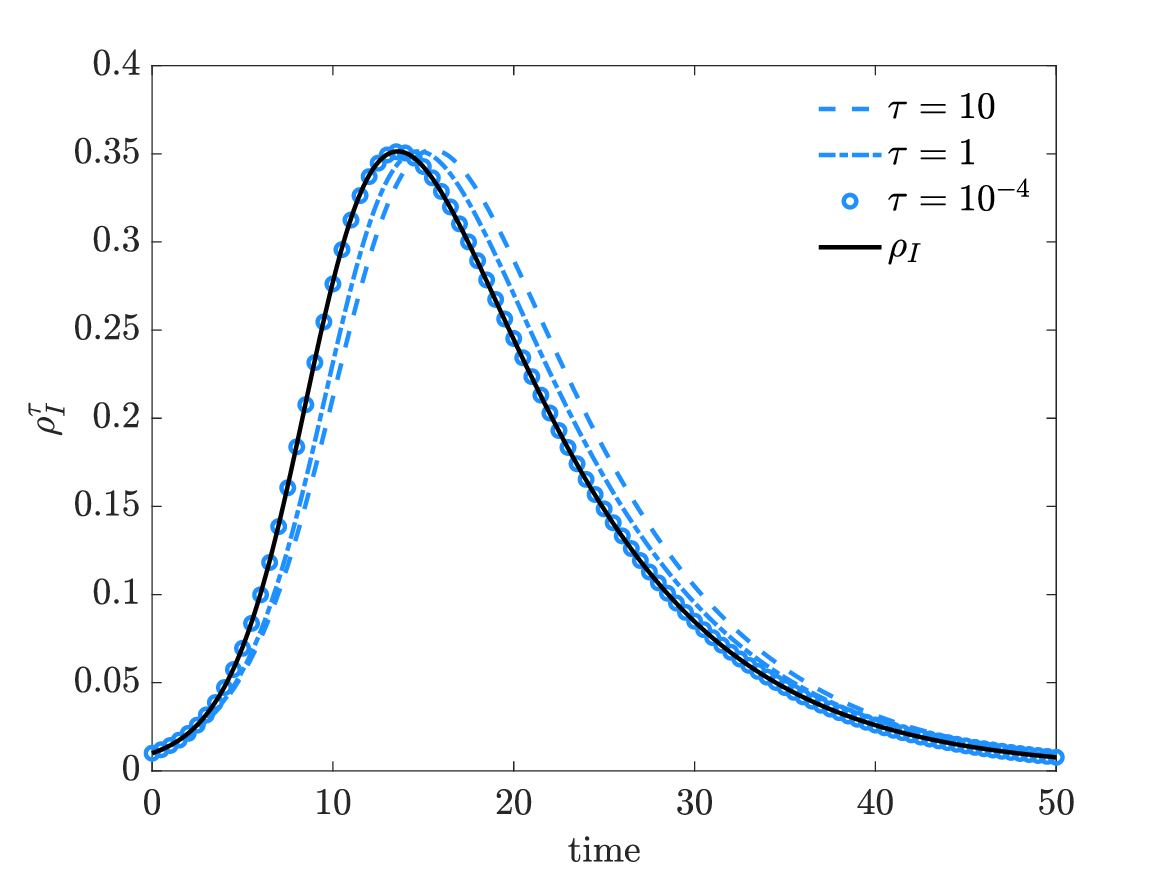}
\includegraphics[scale  = 0.20]{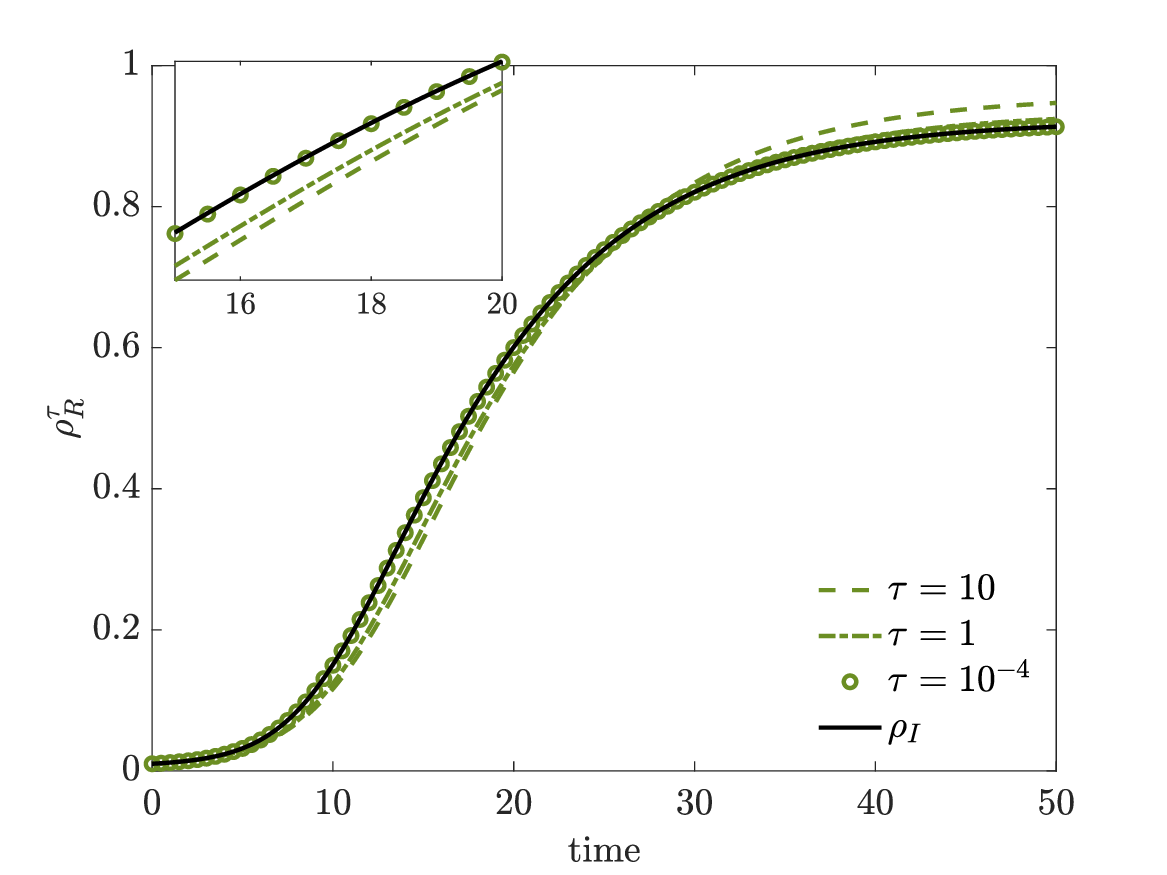} \\
\includegraphics[scale  = 0.20]{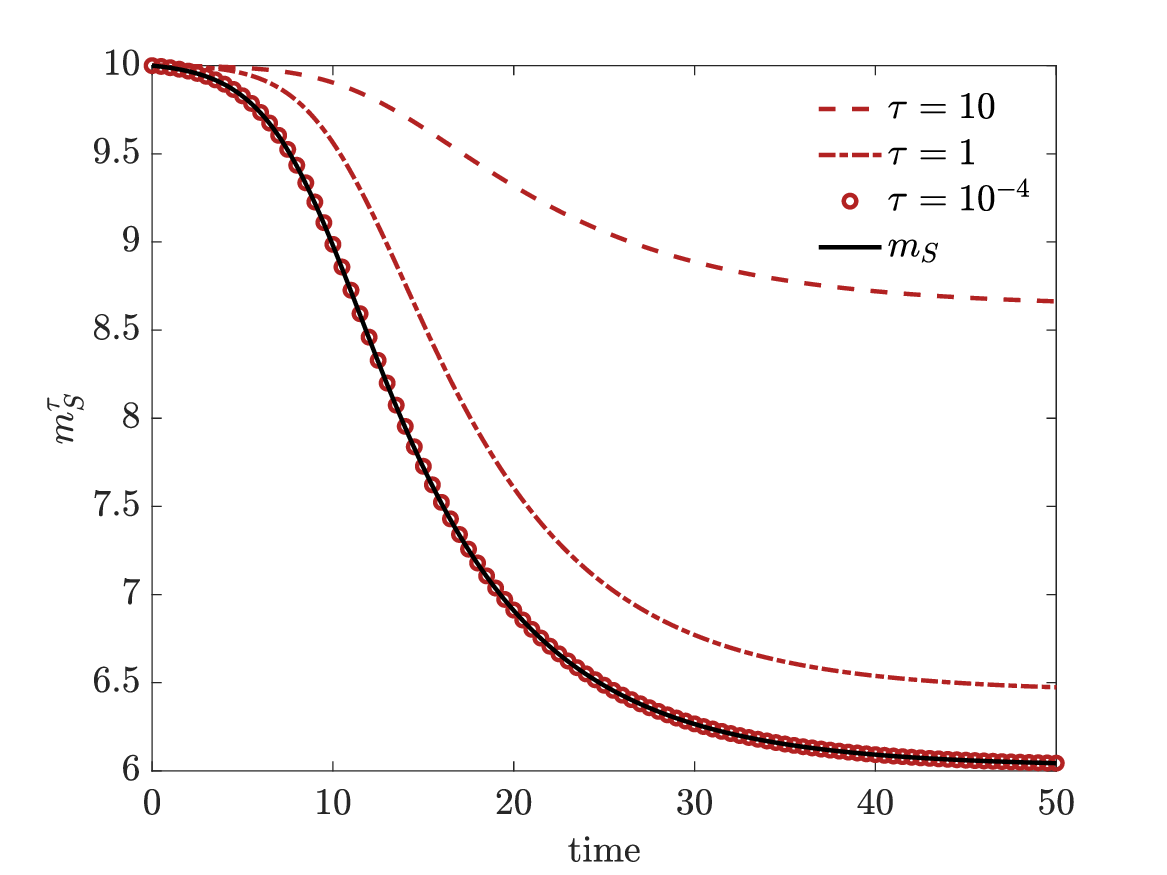}
\includegraphics[scale  = 0.20]{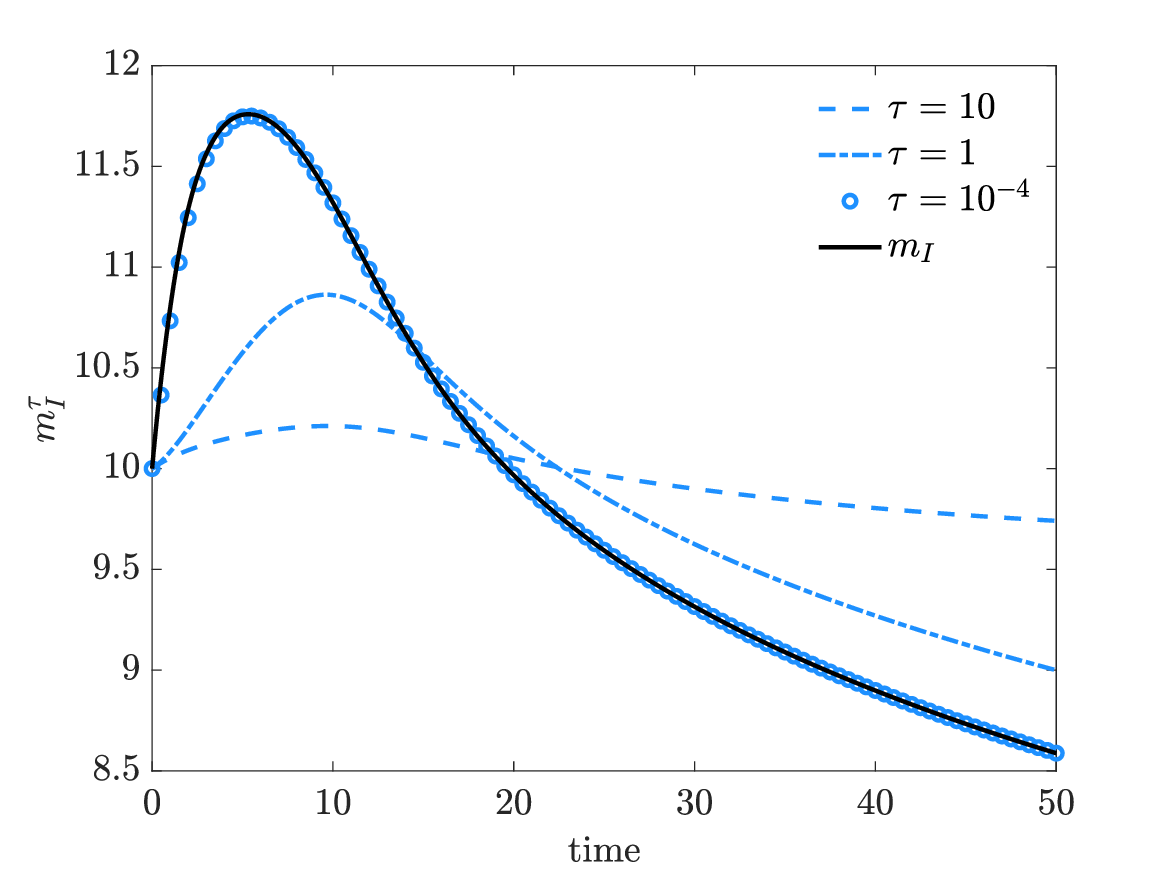}
\includegraphics[scale  = 0.20]{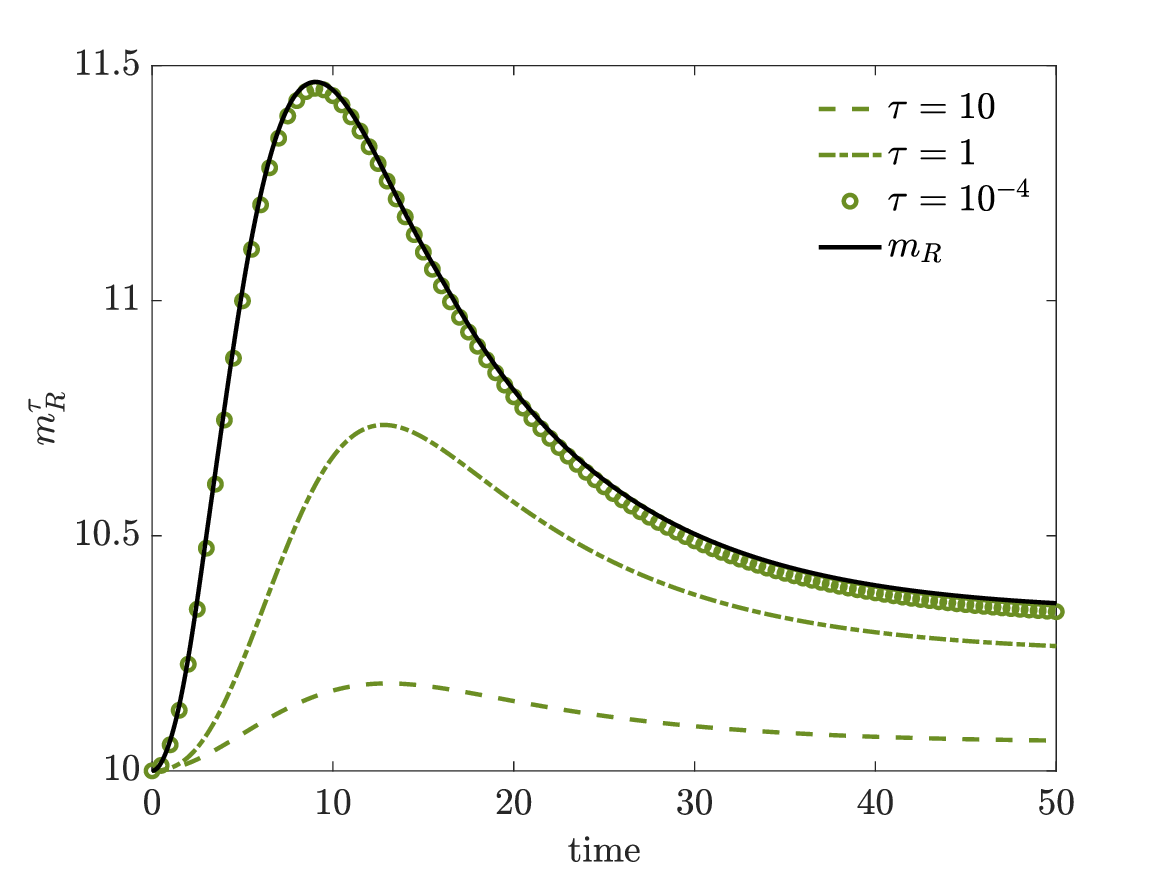} 
\caption{Evolution of mass fractions (top row) and mean number of contacts (bottom row) for the macroscopic model obtained with a Gamma closure approach. We compare the dynamics of $\rho_J(t)$, $m_J(t)$, $J \in \mathcal C$, defined in \eqref{eq:macro_mass}-\eqref{eq:macro_mean_deltapos} with the ones obtained from the approximation of the kinetic model \eqref{eq:kineticSIR} for several values of $\tau = 10^{-4},1,10$. The kinetic densities have been approximated in the interval $[0,50]$ discretized with $N_x = 501$ gridpoints, we considered $\sigma^2 = 0.2$, $\alpha = 1$, $\gamma_I = 1/7$. Initial distribution defined in \eqref{eq:init_dist}. }
\label{fig:deltapos}
\end{figure}

In Figure \ref{fig:deltapos} we may observe that in the limit $\tau\ll 1$ the evolution of the observable quantities reconstructed from the system of kinetic equations collapses to the obtained macroscopic system for a Gamma closure approximation. 

\subsection{Consistency of macroscopic modelling via inverse-Gamma closure approximation }
We compare the evolution of the observable quantities defined in \eqref{eq:obs_num}, where $f_J(x,t)$ is solution to \eqref{eq:kineticSIR} with $\delta = -1$, with $\rho_J(t)$, $m_J(t)$ characterising the macroscopic model obtained with an inverse-Gamma closure approximation, see  \eqref{eq:macro_mass}-\eqref{eq:macro_mean_deltaneg}. To this end, we considered the same parameters introduced in Section \ref{sect:cons_pos}
\begin{figure}
\centering
\includegraphics[scale  = 0.20]{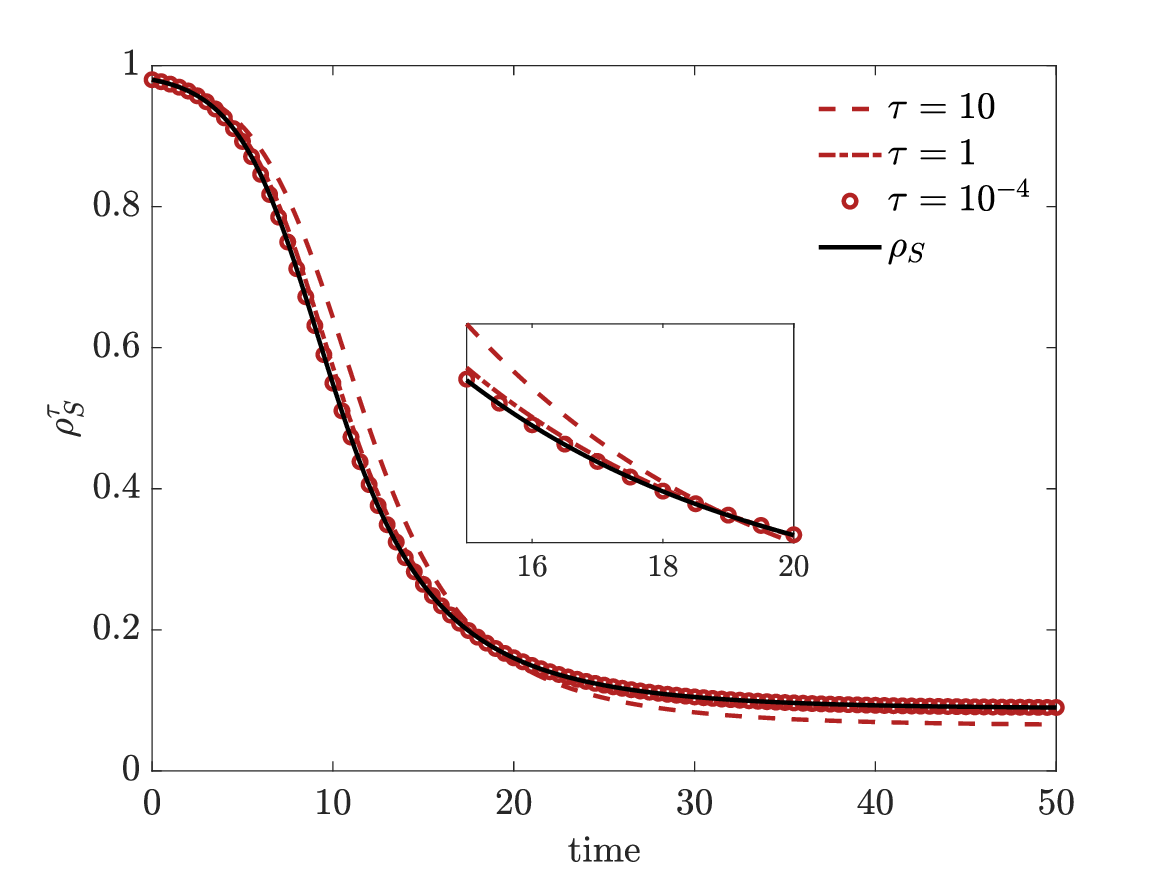}
\includegraphics[scale  = 0.20]{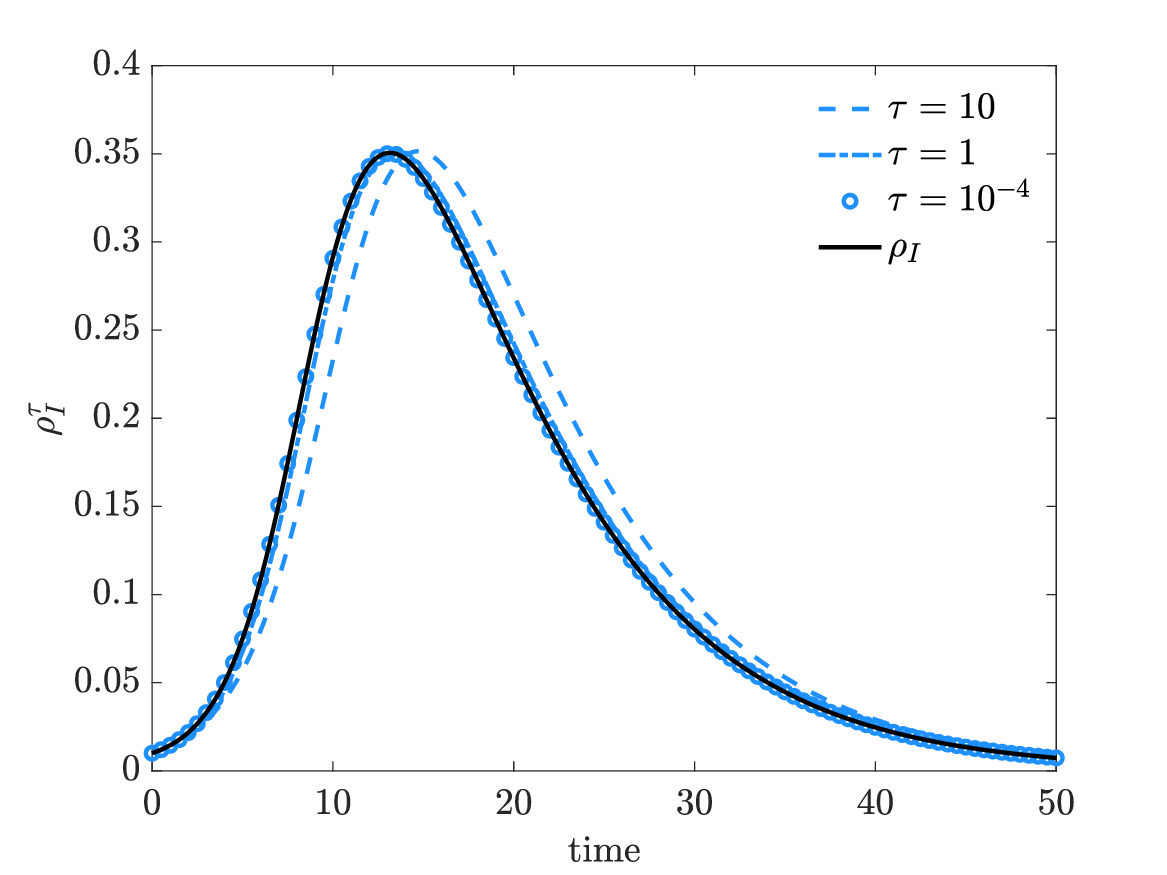}
\includegraphics[scale  = 0.20]{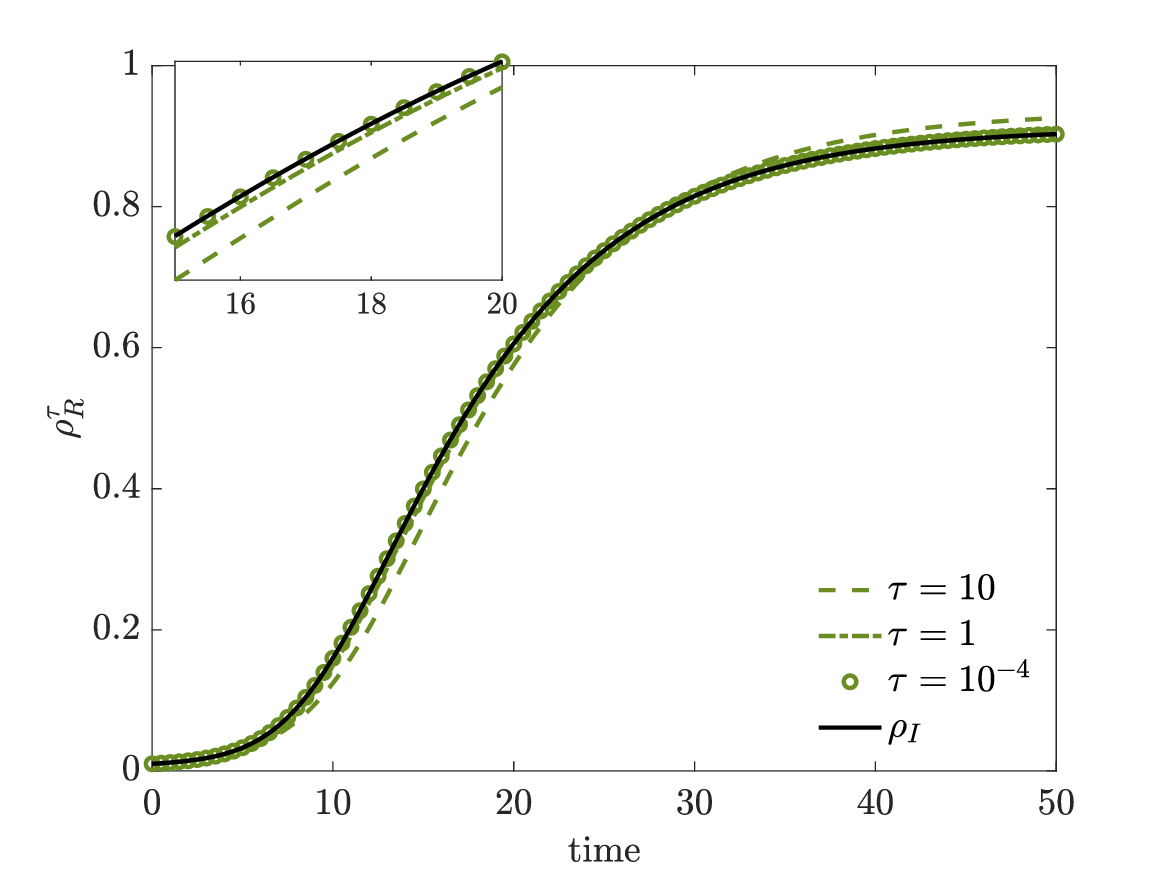} \\
\includegraphics[scale  = 0.20]{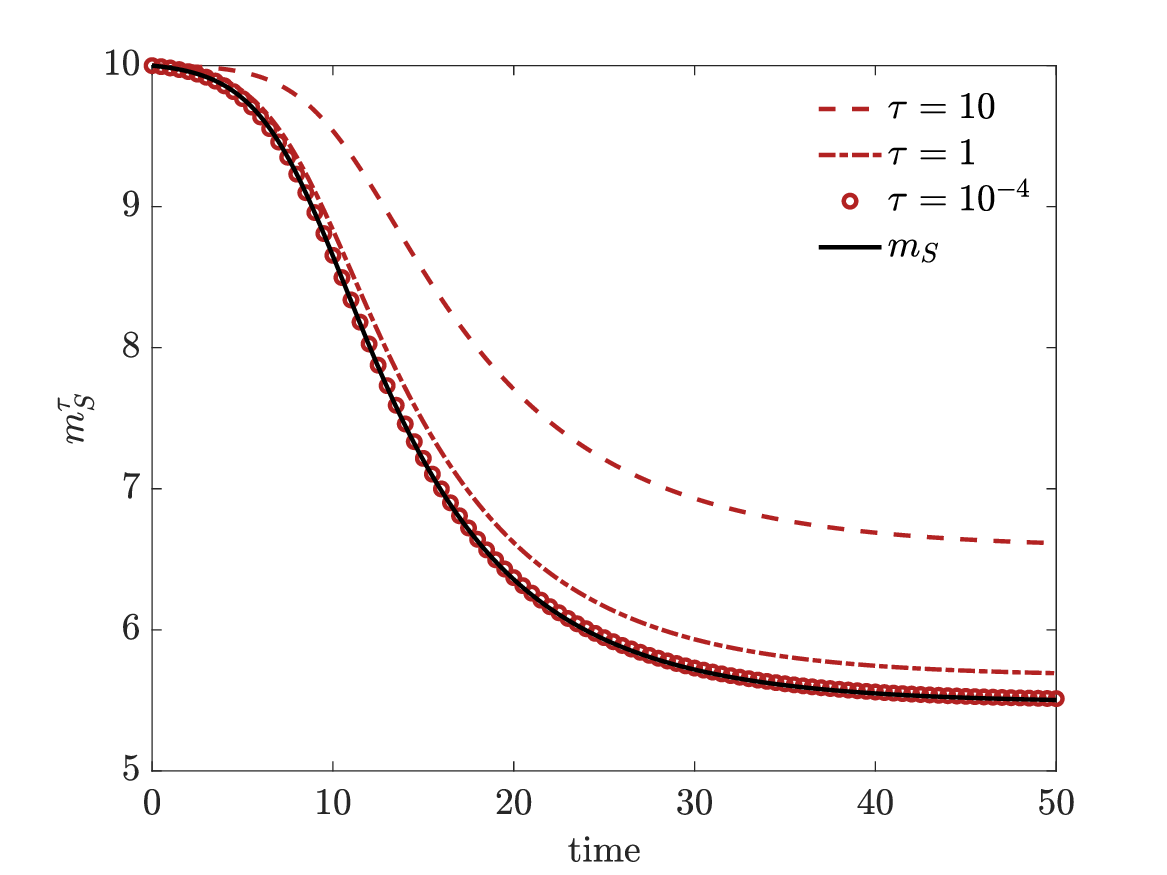}
\includegraphics[scale  = 0.20]{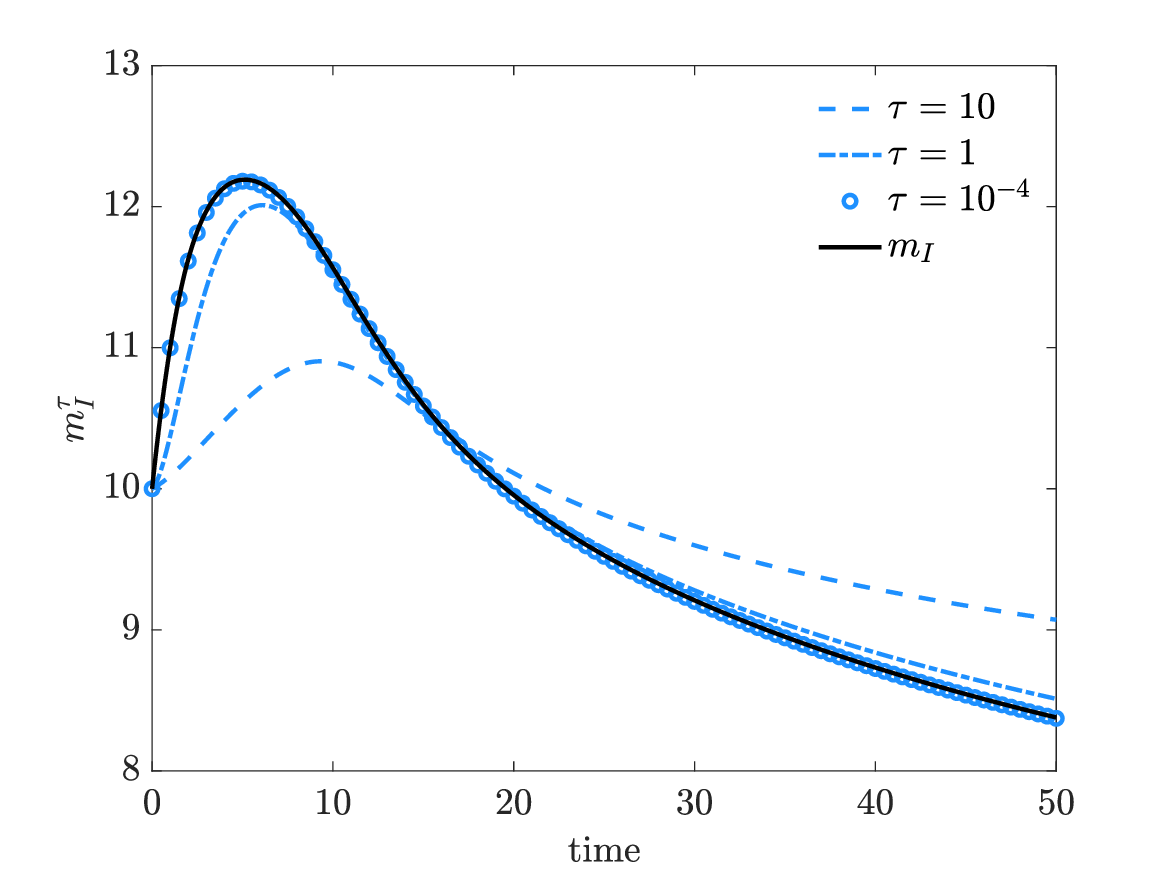}
\includegraphics[scale  = 0.20]{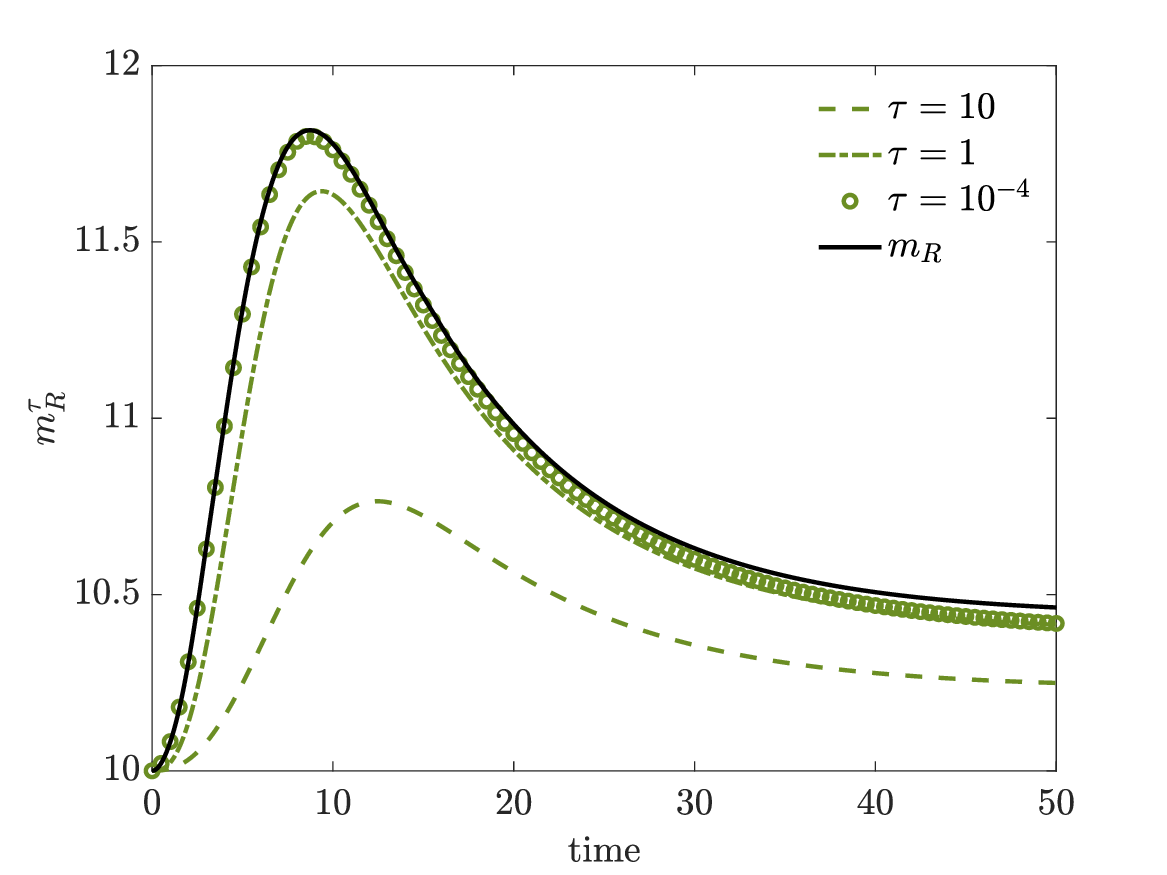} 
\caption{Evolution of mass fractions (top row) and mean number of contacts (bottom row) for the macroscopic model obtained with an inverse-Gamma closure approach. We compare the dynamics of $\rho_J(t)$, $m_J(t)$, $J \in \mathcal C$,
defined in \eqref{eq:macro_mass}-\eqref{eq:macro_mean_deltaneg} with the ones obtained from the approximation of the kinetic model \eqref{eq:kineticSIR} for several values of $\tau = 10^{-4},1,10$. The kinetic densities have been approximated in the interval $[0,50]$ discretized with $N_x = 501$ gridpoints, we considered $\sigma^2 = 0.2$, $\alpha = 1$, $\gamma_I = 1/7$. Initial distribution defined in \eqref{eq:init_dist}.  }
\label{fig:deltaneg}
\end{figure}

In Figure \ref{fig:deltaneg} we may observe that in the limit $\tau\ll 1$ the evolution of the observable quantities reconstructed from the system of kinetic equations collapses to the obtained macroscopic system for an inverse-Gamma closure approximation. 

\section{Concluding remarks and perspectives}
In this work, we discussed a kinetic setting to derive compartmental epidemiological models and that is capable to link  individual agent-based dynamics with macroscopic trends emerging in large populations of agents. By incorporating the dynamics of contact distributions, we bridged the gap between individual behaviors and population-level observable quantities, providing a deeper understanding of how interactions may drive the evolution of an epidemic. The proposed numerical results highlight the robustness of this approach across different regimes, demonstrating its ability to capture the multi-scale nature of epidemiological processes effectively.

A promising extension of the proposed framework is the incorporation of multidimensional social dynamics, such as opinion-type interactions, which could further enrich the modeling of disease transmission by accounting for social factors like public perception and behavior. This direction is currently under deeper investigation, with the potential to capture finer agent-based features that influence epidemiological outcomes at both individual and population levels.

This framework offers significant potential for future research, particularly in exploring the impact of heterogeneous behavioral patterns on disease spread. Future efforts could focus on applying this approach to more complex agent-based dynamics to mimic interactions between different groups. The proposed approach may lead to more accurate predictive models and improved population-specific public health strategies.

\begin{ack}
The author wishes to thank the organisers of the workshop  \emph{Modeling, analysis, and control of multi-agent systems across scales}, held at the Centro di Ricerca Matematica Ennio De Giorgi (Scuola Normale Superiore, Italy), for their kind invitation. 
\end{ack}

\begin{funding}
The work has been written within the activities of GNFM group of INdAM (National Institute of High Mathematics). This work was partially supported by the MUR-PRIN 2020 project (No. 2020JLWP23) “Integrated Mathematical Approaches to Socio–Epidemiological Dynamics” and by European Union - NextGeneration EU.
\end{funding}



\begin{thebibliography}{99}

\bibitem{APZ0}
G. Albi, L. Pareschi, M. Zanella.  Control with uncertain data of socially structured compartmental epidemic models. \emph{J. Math. Biol.}, 82, 63, 2021.

\bibitem{APZ1}
G. Albi, L. Pareschi, M. Zanella. Modelling lockdown measures in epidemic outbreaks using selective socio-economic containment with uncertainty. \emph{Math. Biosci. Eng.}, 18(6): 7161-7190, 2021.

\bibitem{BGL}
C. Bardos, F. Golse, D. Levermore. Fluid dynamic limits of kinetic equations. I. Formal derivations. \emph{J. Stat. Phys.}, 63:323–344, 1991. 

\bibitem{BCGP}
D. Benedetto, E. Caglioti, F. Golse, M. Pulvirenti. Hydrodynamic limits of a Vlasov-Fokker-Planck equation for granular media. \emph{Comm. Math. Sci.}, 2(1):121--136, 2004. 

\bibitem{B_etal}
G. Béraud , S. Kazmercziak, P. Beutels, D. Levy-Bruhl, X. Lenne, N. Mielcarek, Y. Yazdanpanah, P.-Y. Boëlle, N. Hens, B. Dervaux. The French Connection: The First Large Population-Based Contact Survey in France Relevant for the Spread of Infectious Diseases. \emph{PLoS One}, 10(7): e0133203, 2015. 



\bibitem{BZ}
S. Bonandin, M. Zanella. Effects of heterogeneous opinion interactions in many-agent systems for epidemic dynamics. \emph{Netw. Heter. Media}, 19(1): 235-261, 2024.

\bibitem{Catt}
C. Cattuto, W. Van den Broeck, A. Barrat, V. Colizza, J. F.  Pinton, et al. Dynamics of Person-to-Person Interactions from Distributed RFID Sensor Networks. \emph{PLoS ONE}, 5(7): e11596.

\bibitem{Cerc}
C. Cercignani. \emph{The Boltzmann Equation and its Applications}, Springer Series in Applied Mathematical Sciences, vol. 67. Springer-Verlag, New York, NY. 

\bibitem{CPS}
A. Ciallella, M. Pulvirenti, S. Simonella. Kinetic SIR equations and particle limits. \emph{Atti Accad. Naz. Lincei Cl. Sci. Fis. Mat. Natur.}, 32(2): 295--315, 2021. 

\bibitem{Dall}
L. Dall'Amico, J. Kleynhans, L. Gauvin, M. Tizzoni, L. Ozella, M. Makhasi, N. Wolter, B. Language, R. G. Wagner, C. Cohen, S. Tempia, C. Cattuto. Estimating household contact matrices structure from easily collectable metadata. \emph{PLoS One}, 19(3): e0296810, 2024. 

\bibitem{De}
R. Della Marca, N. Loy, A. Tosin. An SIR model with viral load-dependent transmission. \emph{J. Math. Biol.}, 86(4):61, 2023. 

\bibitem{DPTZ} 
G. Dimarco, B. Perthame, G. Toscani, M. Zanella. Kinetic models for epidemic dynamics with social heterogeneity. \emph{J. Math. Biol.}, \textbf{83}, 4, 2021. 

\bibitem{DTZ}
G. Dimarco, G. Toscani, M. Zanella. 
 Optimal control of epidemic spreading in the presence of social heterogeneity. \emph{Phil. Trans. R. Soc. A}, 380:20210160, 2022. 
 

\bibitem{Ferguson}
N. Ferguson, D. Cummings, S. Cauchemez, et al. Strategies for containing an emerging influenza pandemic in Southeast Asia. \emph{Nat.}, 437: 209--214, 2005.

\bibitem{FMZ}
J. Franceschi, A. Medaglia, M. Zanella. On the optimal control of kinetic epidemic models with uncertain social features. \emph{Optim. Contr. Appl. Meth.}, 45(2): 494-522, 2024.

\bibitem{Fumanelli}
L. Fumanelli, M. Ajelli, P. Manfredi, A. Vespignani, S. Merler. Inferring the structure of social contacts from demographic data in the analysis of infectious diseases spread. \emph{PLoS Comput. Biol.},  8(9): e1002673, 2012. 

\bibitem{FPTT}
G. Furioli, A. Pulvirenti, E. Terraneo, G. Toscani. Fokker-Planck equations in the modeling of socio-economic phenomena. \emph{Math. Mod. Meth. Appl. Sci.}, 27(1):115--158, 2017. 

\bibitem{G}
P. Gerlee. The model muddle:  in search of tumor growth laws. \emph{Cancer Res.}, 73(8):2407--2411, 2013. 

\bibitem{Germann}
T. C. Germann, K. Kadau, I. M. Longini Jr., C. A. Macken. Mitigation strategies for pandemic influenza in the United States. \emph{Proc. Natl. Acad. Sci. USA}, 103(15):5935--5940, 2006. 

\bibitem{HP}
A. Hernando, A. Plastino. Scale-invariance underlying the logistic equation and its social applications. \emph{Phys. Lett. A}, 377(3--4):176--180, 2013. 

\bibitem{H}
H. W. Hethcote. The mathematics of infectious diseases. \emph{SIAM Rev.}, 42(4):599 -- 653, 2000. 

\bibitem{LeBL}
C. Le Bris, P.-L. Lions. Existence and uniqueness of solutions to Fokker-Planck type equations with irregular coefficients. \emph{Commun. Partial Differ. Equ.}, 33(7):1272-1317, 2008

\bibitem{LT}
N. Loy, A. Tosin. A viral load-based model for epidemic spread on spatial networks
\emph{Math. Biosci. Eng.}, 18(5):5635--5663, 2021

\bibitem{JMS}
W. Jin, S. W. McCue, M. J. Simpson. Extended logistic growth model for heterogeneous populations. \emph{J. Theor. Biol.}, 445:51--61, 2018. 

\bibitem{MZ}
A. Medaglia, M. Zanella. Kinetic and Macroscopic Epidemic Models in Presence of Multiple Heterogeneous Populations. In: P. Barbante, F. D. Belgiorno, S. Lorenzani, L. Valdettaro, \emph{From Kinetic Theory to Turbulence Modeling}. INdAM 2021. Springer INdAM Series, vol 51.

\bibitem{MA}
S. Merler, M. Ajelli. The role of population heterogeneity and human mobility in the spread of pandemic influenza. \emph{Proc. R. Soc. B}: 277:557--565, 2010. 

\bibitem{Mistry}
D. Mistry, M. Litvinova, A. Pastore y Piontti, M. Chinazzi, L. Fumanelli, M. F. C. Gomes, S. A. Haque, Q.-H. Liu, K. Mu, X. Xiong, M. E. Halloran, I. M. Longini Jr., S. Merler, M. Ajelli, A. Vespignani. Inferring high-resolution human mixing patterns for disease modeling. \emph{Nat. Commun.}, 12, 323, 2021. 

\bibitem{Mossong}
J. Mossong , N. Hens, M. Jit, P. Beutels, K. Auranen, R. Mikolajczyk, M. Massari, S. Salmaso, G. Scalia Tomba, J. Wallinga, J. Heijne, M. Sadkowska-Todys, M. Rosinska, W. J. Edmunds.  Social contacts and mixing patterns relevant to the spread of infectious diseases. \emph{PLoS Med.}, 5(3): e74, 2008.

\bibitem{PT}
L. Pareschi, G. Toscani. \emph{Interacting Multiagent Systems: Kinetic Equations and Monte Carlo Methods}. Oxford University Press, 2013. 
\bibitem{PZ}
L. Pareschi, M. Zanella.  Structure preserving schemes for nonlinear Fokker-Planck equations and applications. \emph{J. Sci. Comput.}, 74(3): 1575-1600, 2018.

\bibitem{Pav}
G. A. Pavliotis. \emph{Stochastic Processes and Applications: Diffusion Proceses, the Fokker-PLanck and Langevin Equations}. Springer New York, NY, 2014. 

\bibitem{Pol}
P. Poletti, B. Caprile, M. Ajelli, A. Pugliese, S. Merler. Spontaneous behavioural changes in response to epidemics. \emph{J. Theor. Biol.}, 260(1):31--40, 2009. 

\bibitem{Risken}
H. Risken. \emph{The Fokker-Planck Equation: Methods of Solution and Applications}. Springer Berlin, Heidelberg, 1996. 

\bibitem{RBKW}
I. A. Rodriguez-Brenes, N. L. Komarova, D. Wodarz. Tumor growth dynamics: insights into evolutionary processes. \emph{Trends Ecol. Evol.}, 28(10):597--604, 2013. 

\bibitem{TPZ}
G. Toscani, L. Preziosi, M. Zanella. Control of tumor growth distributions through kinetic methods. \emph{J. Theoret. Biol.}, 514: 110579, 2021. 

\bibitem{Toscani_Gamma}
G. Toscani. Entropy-type inequalities for generalized Gamma densities. \emph{Ric. Mat.}, 70:35--50,2021. 

\bibitem{WBE}
G. B. West, J. H. Brown, B. J. Enquist. A general model for ontogenetic growth. \emph{Nat.},  413: 628--631, 2001. 

\bibitem{vonB}
L. von Bertalaffy. Quantitative laws in metabolism and growth. \emph{The Quarterly Review of Biology}, 32(3):217--231, 1957. 

\bibitem{Z}
M. Zanella. Kinetic models for epidemic dynamics in the presence of opinion polarization. \emph{Bull. Math. Biol.}, 85:36, 2023.

\bibitem{Z_etal}
M. Zanella, C. Bardelli, G. Dimarco, S. Deandrea, P. Perotti, M. Azzi, S. Figini, G. Toscani. A data-driven epidemic model with social structure for understanding the COVID-19 infection on a heavily affected Italian Province. \emph{Math. Mod. Meth. Appl. Sci.}, 31(12):2533--2570, 2021.

\bibitem{Zha}
J. Zhang et al. Changes in contact patterns shape the dynamics of the COVID-19 outbreak in China. \emph{Science}, 368:1481--1486, 2020. 


\bibitem{K}
S. E. Kingsland. \emph{Modeling Nature}. The University of Chicago Press, Chicago and London, 1985. 


\end{thebibliography}
\end{document}